\documentclass[]{spie}  

\usepackage{amsmath,amsfonts,amssymb}
\usepackage{graphicx}
\usepackage[colorlinks=true, allcolors=blue]{hyperref}

\usepackage{dcolumn}
\usepackage{bm}
\usepackage{wrapfig}
\usepackage{verbatim}
\usepackage{xcolor} 
\usepackage{soul}
\usepackage{adjustbox}
\usepackage{subcaption}
\usepackage{hyperref}

\usepackage{caption}
\usepackage{subcaption}
\usepackage{mathtools}

\usepackage{hyperref}
\hypersetup{
        colorlinks=true,       
        linkcolor=blue,          
        citecolor=blue,        
        filecolor=magenta,      
        urlcolor=blue
        }

\newlength{\wdth}

\title{Electron Beam Profiling via Rydberg Electromagnetically Induced
Transparency in Rubidium Vapor with Crossed Laser beams}

\author[a]{Rob Behary}
\author[a]{Kevin Su}
\author[a]{Nicolas DeStefano}
\author[a]{Jennifer Tsai}
\author[a]{Todd Averett}
\author[b]{Alexandre Camsonne}
\author[b]{Shukui Zhang}
\author[c]{Charles T. Fancher}
\author[c]{Neel Malvania}
\author[a]{Seth Aubin}
\author[a]{Eugeniy E. Mikhailov}
\author[a]{Irina Novikova}
\affil[a]{Department of Physics, William \& Mary, 300 Ukrop Way, Williamsburg, VA 23185, USA}
\affil[b]{Thomas Jefferson National Accelerator Facility, 12000 Jefferson Avenue, Newport News, VA 23606, USA}
\affil[c]{The MITRE Corporation, McLean, VA 22102, USA}

\authorinfo{Further author information: 
\\Rob Behary: E-mail: rbehary@wm.edu
\\Irina Novikova: E-mail: inovikova@physics.wm.edu
}

\begin{document} 
\maketitle

\begin{abstract}
We present an all-optical detection approach to determine the position and spatial profile of an electron beam based on quantum properties of alkali metal atoms. To measure the electric field, produced by an electron beam, we excite thermal rubidium atoms to a highly excited Rydberg state via a two-photon ladder transition and detect Stark shifts of Rydberg states by monitoring frequencies of the corresponding electromagnetically induced transparency (EIT) transmission peaks. We addressed several technical challenges in this approach. First, we use crossed laser beams to obtain spatial information about the electron beam position and geometry. Second, by pulsing the electron beam and using phase-sensitive optical detection, we separate the true electron beam electric signature from the parasitic electric fields due to photoelectric charges on the windows. Finally, we use a principle component analysis to further improve signal quality. We test this method to detect the current and to reconstruct a 2D profile of a 20 keV electron beam with currents ranging from 25~$\mu A$ - 100~$\mu A$. While this technique provides less spatial resolution than fluorescence-based measurements, thanks to their speed and limited optical access requirements it can be useful for real-time non-invasive diagnostics of charged particle beams at accelerator facilities.
\end{abstract}

\keywords{electron beam, Rydberg atoms, Electromagnetically Iduced Transparency, atomic vapor, optical diagnostics}

\section{Introduction}

\begin{figure*}
    \center
    \includegraphics[width=0.9\textwidth]{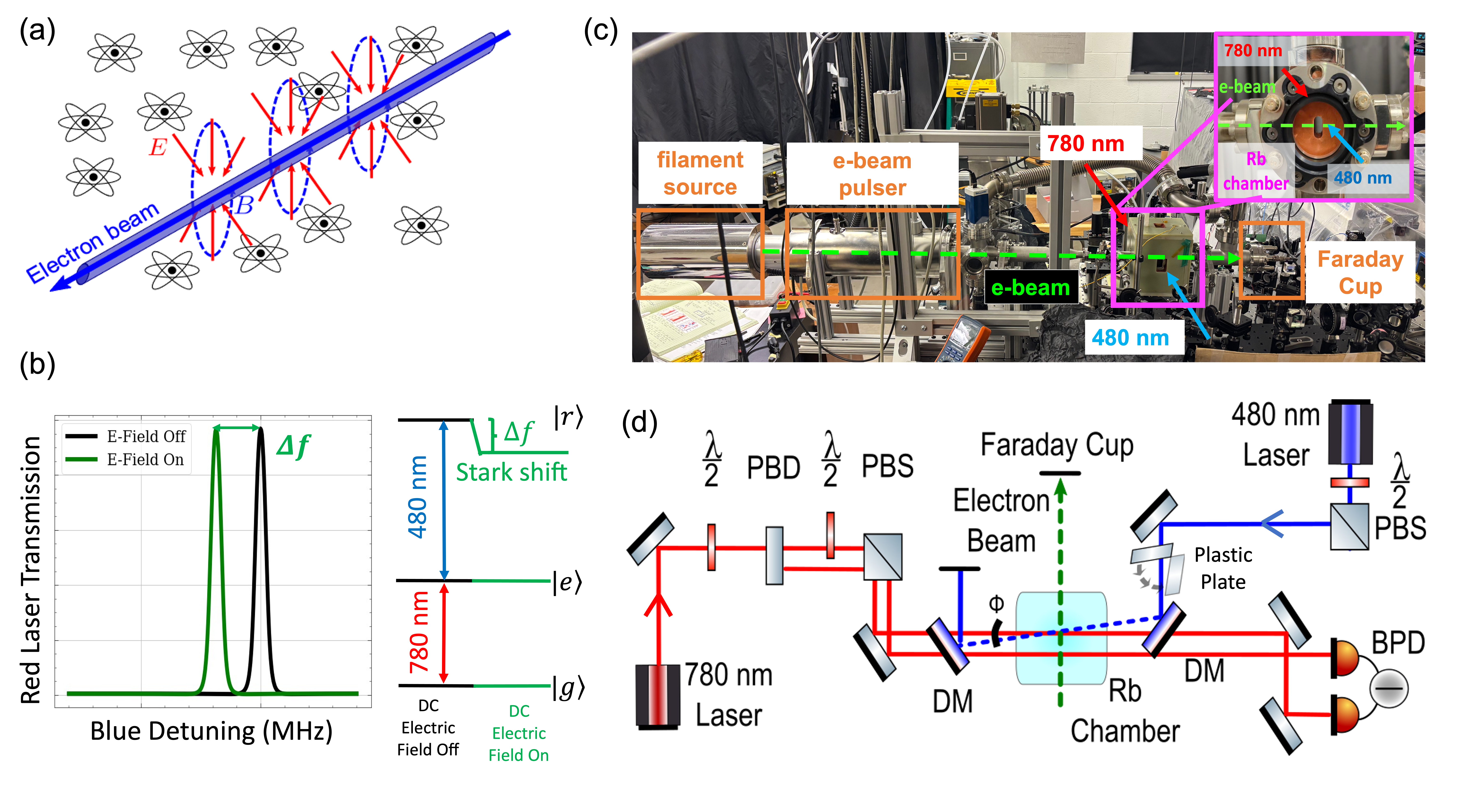}
    \caption{An overview of the experimental apparatus and detection method.
    (a) Our detection medium is a low density rubidium vapor that surrounds an electron beam. As the electron beam propagates it produces an electric and magnetic field that perturbs the atoms.
    (b) A simplified detection scheme is monitoring the dc Stark shifts to the Rydberg atom energy level. The change in frequency monitored through EIT can be used to know the strength of the applied electric field.
    (c) Experimental setup to show the electron beam path and Rb chamber.
    (d) Simplified optical setup. We show that instead of the standard counter-propagating Rydberg EIT detection method, we can cross the beams at an angle within the Rb chamber to gain spatial resolution within the cell.}
    \label{fig:apparatus}
\end{figure*}

Atoms excited to a high principle quantum number (n $\geq$ 20) are known as Rydberg atoms~\cite{Gallagher_1994}.
These atoms have a large polarizability~\cite{ARC,Gallagher_1994} which makes them an attractive platform for a variety of applications related to electric field sensing~\cite{schlossberger2024rydberg}.
For example, they are candidates for a voltage standard~\cite{HollowayJAP2017}, used for rf field receivers~\cite{FrancherIEEE2021,Simons:18,RFSensingReview}, and implemented in THz imaging applications~\cite{Downes_2023,PhysRevX.10.011027}.
Coherent optical resonances, in particular electromagnetically induced transparency (EIT), is  beneficial for Rydberg state spectroscopy. 
For such experiments Rb atoms typically interact with two counter-propagating lasers (780~nm infrared and 480 nm blue).
Since external electric fields shift the Rydberg state sublevels due to the Stark effect, the strength of the applied electric field can be deduced from changes reflected in changes in the EIT transmission spectra, where the frequency position of the shifted EIT peaks allow for accurate electric field measurements~\cite{Scully,lukema,thesis2011,OptimizingEITVapor,schlossberger2024rydberg}.
%
However, such experimental configurations are challenging for spatially inhomogeneous electric fields, since reconstruction of its distribution requires measuring Stark shifts at different locations, and EIT transmission integrates the signal along the beam propagation direction. Previous experiments~\cite{hollowayAPL2014,Ma:20,anderson2017} explored one spatial dimension, and required readout of an integrated signal along the entire vapor cell. In this case the spatial variations of the electric field (in both dc and rf regimes) are reflected in the distortions of the EIT lineshapes~\cite{hollowayAPL2014,NonUniformRFFields,anderson2017,PhysRevApplied.18.024001}.Two-dimensional field reconstruction can be efficiently performed by imaging the EIT-induced changes in fluorescence~\cite{SchlossbergerOL25,BeharyPRR2025}. However, this method require additional optical access to the measurement volume for fluorescence detection, and may have a limited detection bandwidth due to camera speed.


Here we explore an alternative approach for a 2D spatial electric field mapping that is based on the transmission  lineshape analysis for non-collinear EIT geometry. 
We apply this method to measure the position and current of an electron beam (\textit{e}-beam), as an alternative to magnetic field-based imaging~\cite{DeStefano_APL2024} or fluorescence imaging~\cite{BeharyPRR2025}. The general concept of the  detection is illustrated in Fig.~\ref{fig:apparatus}(a and b). A collimated \textit{e}-beam is practically a wire of charge that produces electric and magnetic fields, as shown in Fig.~\ref{fig:apparatus}(a). If the volume around the \textit{e}-beam contains some Rb vapor, each Rb atom can sense its local electric field when exited into a Rydberg state. Electric fields lift the degeneracy of a Rydberg state depending on its quantum number $m_J$~\cite{hollowayAPL2014,schlossberger2024rydberg,PhysRevApplied.18.024001}:
\begin{equation}
\label{eq:stark-shift}
    h\Delta f_{|m_J|}(E) = -\frac{1}{2}\alpha_{|m_J|}E^2.
\end{equation}
Where $\Delta f_{|m_J|}(E)$ is the dc Stark shift, $h$ is Planck's constant, $\alpha_{|m_J|}$ is the polarizability of each $m_J$ sublevel of the Rydberg state, and $E$ is the magnitude of the electric field. 
The above equation approximates the shifts of small electric fields. A more accurate description of these shifts are obtained numerically by the Alkali Rydberg calculator (ARC)~\cite{ARC}.
Analysis of the Stark shifted EIT spectra provides information about electric field experienced by Rydberg atoms. In this experiment instead of using traditional counter-propagating optical fields, we introduced a small angle between them to reduce the E-field sensing volume which is constrained by the beam overlapping region. A drawback of this method is increased Doppler broadening of EIT resonances~\cite{Su2024} leading to lower electric field sensitivity. Nevertheless, by rastering the 2D detection space, we are able to reconstruct the profile of the \textit{e}-beam with sub-mm precision. 


\section{Apparatus/Detection Method}

Our experimental apparatus is shown in Fig~\ref{fig:apparatus}(c) (see reference~\cite{BeharyPRR2025} for more details).
We use a Staib Instruments filament source that emits a collimated 20~keV \textit{e}-beam with full width at half maxima (FWHM) width $\approx 1$~mm in continuous or pulsed regime. The pulser can turn the \textit{e}-beam on and off at a rate up to 1~MHz, as well as steer the beam horizontally and vertically within a small solid angle.
The \textit{e}-beam travels down a vacuum system, crosses a Rb interaction chamber, and at the end hits a Faraday cup where we monitor the \textit{e}-beam current. The  Rb interaction chamber  is a six-way cross with three optical windows. A Rb metal ampule placed in the bottom of the chamber, and the whole interaction chamber is placed inside a hot-air oven to control the saturated vapor density of Rb. Narrow (8~mm diameter) tubes before and after the chamber effectively prevent Rb spreading to the rest of the vacuum system. The laser beams cross the chamber perpendicularly to the electron beam, as shown in Fig.\ref{fig:apparatus}(d).  
Independent measurements of the \textit{e}-beam parameters are done using impact florescence, in which we increase rubidium density and detect its fluorescence caused by collisions with electrons. For these measurements we use a ccd camera with long exposure time ($\approx 30$~s), mounted on top of the interaction chamber.
Impact florescence provides  reference values for the \textit{e}-beam width at the location of Rydberg atoms, and allows calibration of the \textit{e}-beam steering.


A simplified optical setup is shown in Fig.~\ref{fig:apparatus}(d).
The blue laser power is fixed at 60 mW and has a beam FWHM size of $\approx$ 0.2 mm.
The 780~nm red laser is tuned and locked to the ${}^{85}$Rb 5S${}_{1/2}$ (F=3) $\rightarrow$ 5P${}_{3/2}$ (F'=2,3 crossover) resonance using a saturation absorption spectroscopy reference. 
The red laser power is fixed at 60 $\mu$W and is split with a polarizing beam displacer into two identically polarized beams both with a FWHM of $\approx$ 0.2 mm. Only one of these beams  overlaps with the blue beam, producing EIT resonances, and the other provides a reference.
The power difference between the beams allows accurate detection of even weak EIT resonances without large background. For even more accurate detection the output of EIT detector is sent to a lock-in amplifier that is synced with the pulsing of the electron beam at 5~kHz. We found that this pulse speed allowed for minimal charging of the chamber and strong response from the \textit{e}-beam.

In the Rb interaction chamber, the red and blue beams are crossed at an angle $\phi \approx$ 7${}^{\circ}$, creating a sensing region of $\approx 5$ mm (see Fig.~\ref{fig:raster_motion}) that is smaller then the smallest Rb chamber dimension (3~cm). The size of this region  in $z$ and $y$ dimensions is set by the laser beam cross-section. 
While better spatial resolution may be achieved with larger angle between the beams $\phi$, we here are limited by
by opening size in the copper plates placed within the Rb chamber to mitigate charging on the glass windows. Also, larger $\phi$ causes rapid broadening of EIT peaks~\cite{Su2024}, and we wanted to keep the EIT resonance width to below 100~MHz to remain sensitive to the electric field.

\section{Experimental results}

\begin{figure}
    \center
    \includegraphics[width=0.7\linewidth]{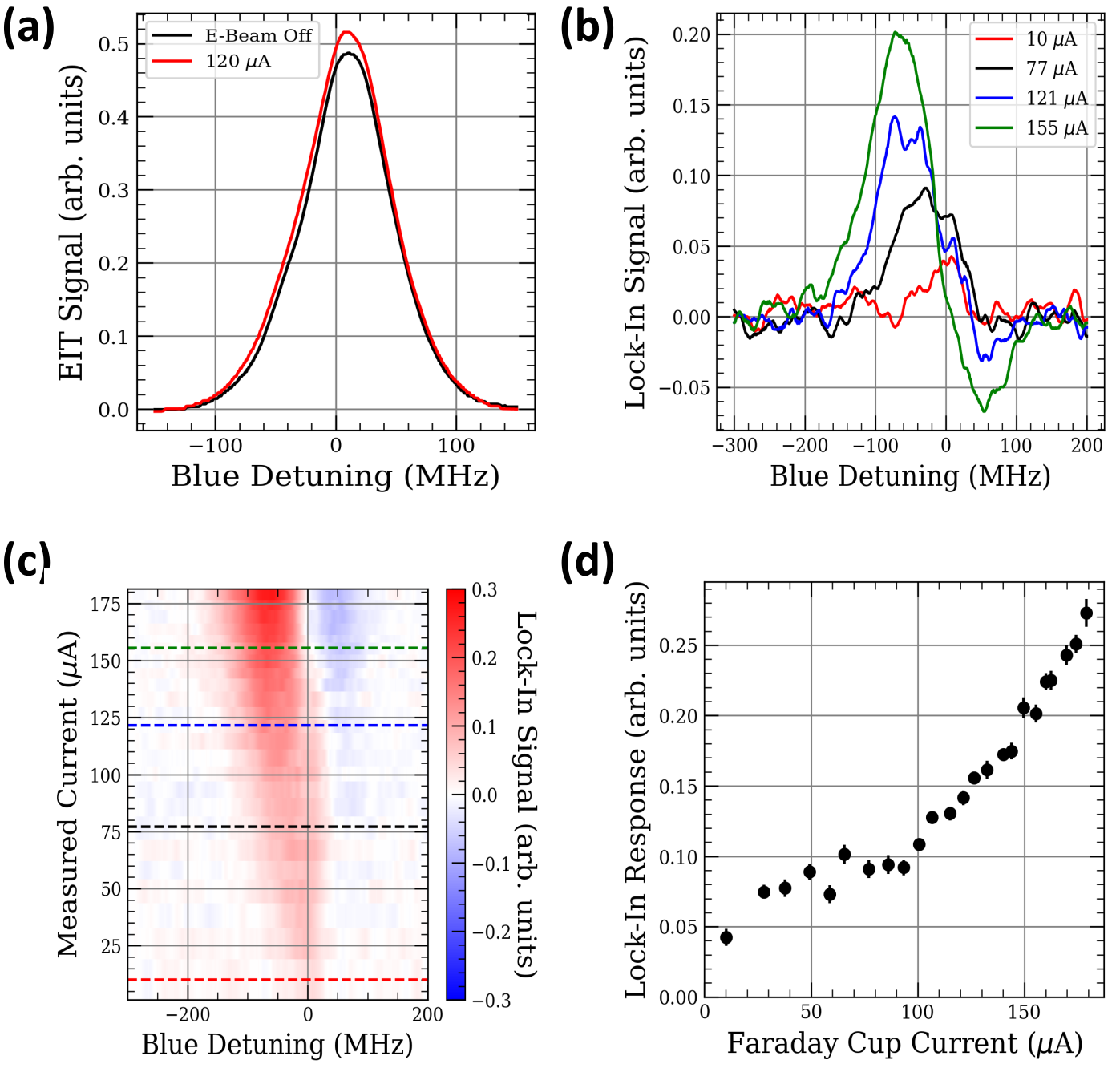}
    \caption{
    (a) EIT spectra for the case of the \textit{e}-beam off and on with 120~$\mu$A total current.
    (b) Lock-in signal due to pulsing of the \textit{e}-beam for different currents.
    (c) The heat map is the lock-in signal strength vs the blue laser detuning and the \textit{e}-beam current. Dashed lines corresponds to  the location of the lock-in curves shown (b).
    (d) Lock-in signal maximum value vs the \textit{e}-beam current.
}
    \label{fig:lock-in-example}
\end{figure}

To measure the EIT spectra we lock the red laser and sweep the frequency of the blue laser around the 58D Rydberg resonance.
Fig.~\ref{fig:lock-in-example}(a) shows an example of our recorded EIT spectra due to the presence of the \textit{e}-beam.
To overlap the \textit{e}-beam and the lasers, we use polarization rotation methods to align the red laser with the \textit{e}-beam~\cite{DeStefano_APL2024}, and then move the blue laser until the spectra difference is maximized between \textit{e}-beam on and off cases.
The spectrum frequency axis is calibrated with a separate EIT reference cell by observing a known hyperfine splitting of the 58D${}_{5/2}$ and 58D${}_{3/2}$ sub-levels~\cite{ARC}.
As shown Fig.~\ref{fig:lock-in-example}(a), the EIT spectra have FWHM of about 100~MHz due to the laser beams crossing at non zero angle $\phi$, this is much broader then approximately 10~MHz EIT width in a counter propagation geometry. The electric field produced by the \textit{e}-beam is inhomogeneous in the sensing region and not strong enough to visibly split the 100~MHz wide spectral line, like in the ideal case in Fig.~\ref{fig:apparatus}(b). Consequently,the \textit{e}-beam produces some reshaping of the EIT spectrum, shown for comparison for the \textit{e}-beam on and off  in Fig.~\ref{fig:lock-in-example}(a). For small \textit{e}-beam current this reshaping is very small, and can be further distorted by parasitic electric fields inside the chamber. 

The solution is to pulse the \textit{e}-beam on and off at 5~kHz modulation frequency and performed lock-in detection. This distills and amplifies the electron contribution.
Fig.~\ref{fig:lock-in-example}(b) shows a much higher contrast of the lock-in signal with more easily resolvable features within the EIT lineshape.
Fig.~\ref{fig:lock-in-example}(c) shows a heat map of such signals as a function of the \textit{e}-beam current ($y$-axis)  and the blue laser detuning ($x$-axis). There is a clear growth of maximum lock-in signal with the \textit{e}-beam current. We can see the location of the lock-in signal maximum shifts to the negative blue detuning as the electron current, and correspondingly its electric field, increases, in good agreement with the theoretical expectation of the negative Stark shift for $m_J=5/2$ level. We plot the maximum value of the lock-in signal
as a function of the \textit{e}-beam current in Fig.~\ref{fig:lock-in-example}(d). This dependence is monotonic, even if not strictly linear, 
and it can be used as a calibration curve for the value of the current. Thus, our scheme provides a non invasive way to measure the \textit{e}-beam current.

\begin{figure}
    \center
    \includegraphics[width=0.7\linewidth]{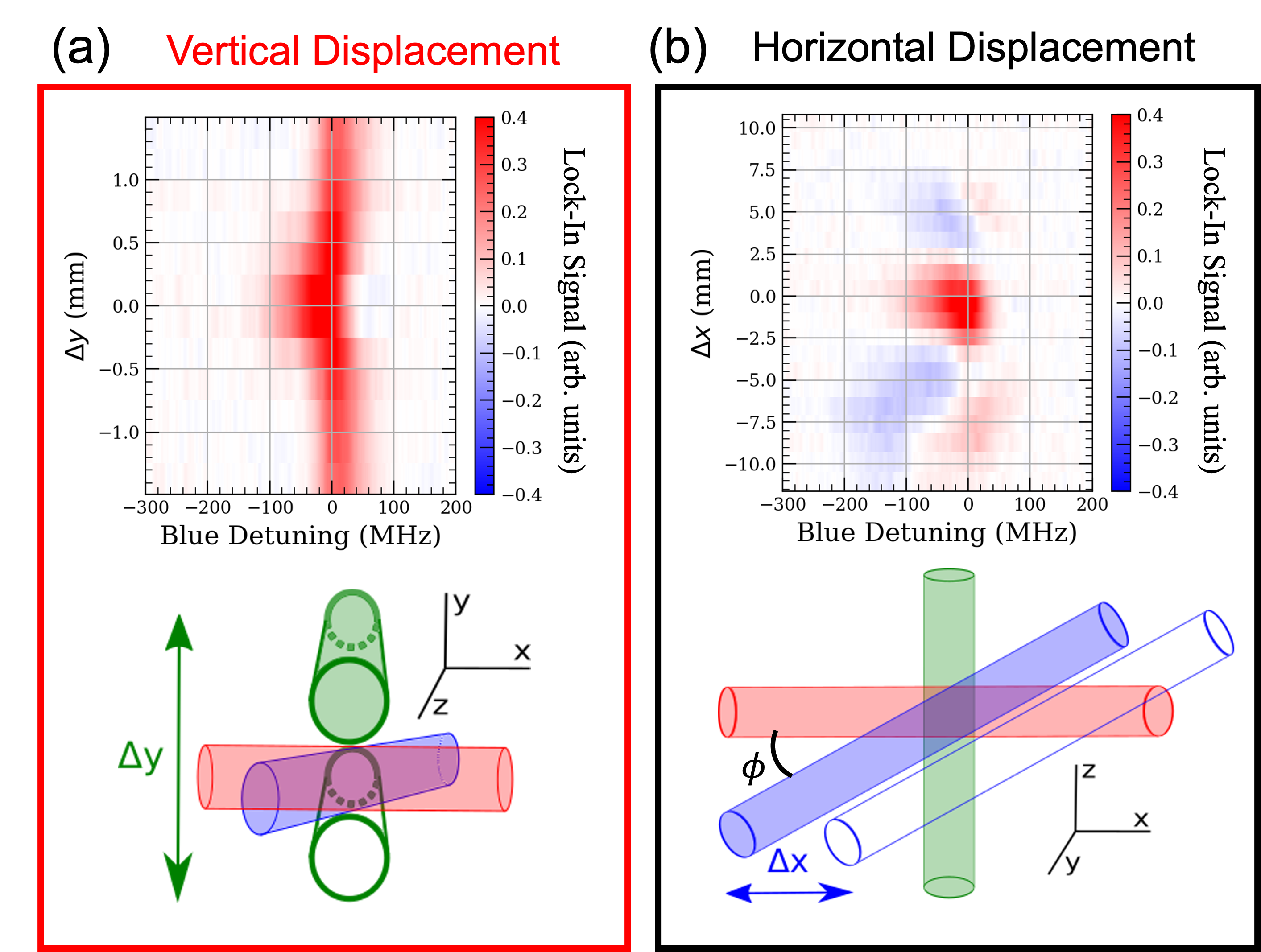}
    \caption{Results (top row) and geometrical arrangements (bottom row) of the electric field measurements in the horizontal and vertical directions. In both plots the red/blue cylinders represent corresponding laser beams, and the green cylinder is the \textit{e}-beam.
    (a) The measured Stark shifts of the EIT resonances as the electron beam is moved vertically through the crossed laser beam region. 
    (b) The lock-in signal modifications as the the blue laser beam is moved horizontally with a thick plastic plate. The central strong red spot corresponds to the location of the electron beam.
	The \textit{e}-beam current was set to 108~$\mu$A.
    \label{fig:raster_motion}
   }
\end{figure}

To gain spatial information about \textit{e}-beam profile, we raster relative position of the \textit{e}-beam and the sensing
region in $x-y$ plane perpendicular to the \textit{e}-beam propagation direction ($z$). To change vertical (along $y$) relative position, we deflect
the \textit{e}-beam relative to the lasers with the pulser \textit{e}-beam steering controls, see Fig.~\ref{fig:raster_motion}(a).
This is not ideal, since motion of the \textit{e}-beam could change the electric field environment in the chamber due to stray
charge deposition. But this is easier, since we do not have to translate two  laser beams in sync to each other. The heat map of
lock-in signals in Fig.~\ref{fig:raster_motion}(a) shows the presence of the feature at $\pm0.5$~mm range around 0 displacement.
Since the spatial resolution is defined by the convolution of the laser beams and the \textit{e}-beam diameters, in the $y$-direction
it is limited by the \textit{e}-beam diameter of $1$~mm which matches the size of the spectral feature. For large $y$-displacements, we would expect the lock-in signal to be zero, but the largest displacements of $\pm1.5$~mm are not large enough to move the \textit{e}-beam
outside of the EIT sensing region. Nevertheless the lock-in signal drops with the increase of displacement.

Since the sensing region size along $x$ is about 5~mm and we need to raster at least factor of two larger range,
we cannot use the \textit{e}-beam steering controls in this direction. Otherwise
the \textit{e}-beam would be clipped by its entrance aperture of about 8~mm diameter. Thus we change the $x$ (horizontal)
relative position $\Delta x$
by translation of the blue beam with a tilt of a thick acrylic plastic plate (see Fig.~\ref{fig:apparatus}d). The tilt angle $\theta$ is geometrically connected to the beam translation $d$ and $\Delta x$ by the following equations
\begin{subequations}
    \begin{equation}
        d = t[\tan\theta - \tan( \arcsin[\frac{\sin\theta}{n_p}])]\cos\theta
    \end{equation}
    \begin{equation}
        \Delta x = d/\sin\phi.
    \end{equation}
\end{subequations}
Where $t=1.2$~cm is the thickness of the  plastic plate, $n_p=1.5$ is the index of refraction of the plate, and $\phi\approx7{}^{\circ}$ is the angle of intersection between the red and blue lasers.
Fig.~\ref{fig:raster_motion}(b) shows the resulting lock-in signal  heat map vs the horizontal motion.
From its asymmetric shape with respect to zero displacement location and non vanishing signal at large displacements,
it is apparent that there is an electric field gradient generated
by the pulsing \textit{e}-beam within the Rb chamber. This can be attributed to the modulation of charges
on the glass optical window ports via Rb ionization or any number of charging effects
that we do not have the infrastructure to mitigate in our Rb chamber.
However, there is a strong  feature near zero displacement that is due to the \textit{e}-beam itself.

\begin{figure}
    \center
    \includegraphics[width=1\linewidth]{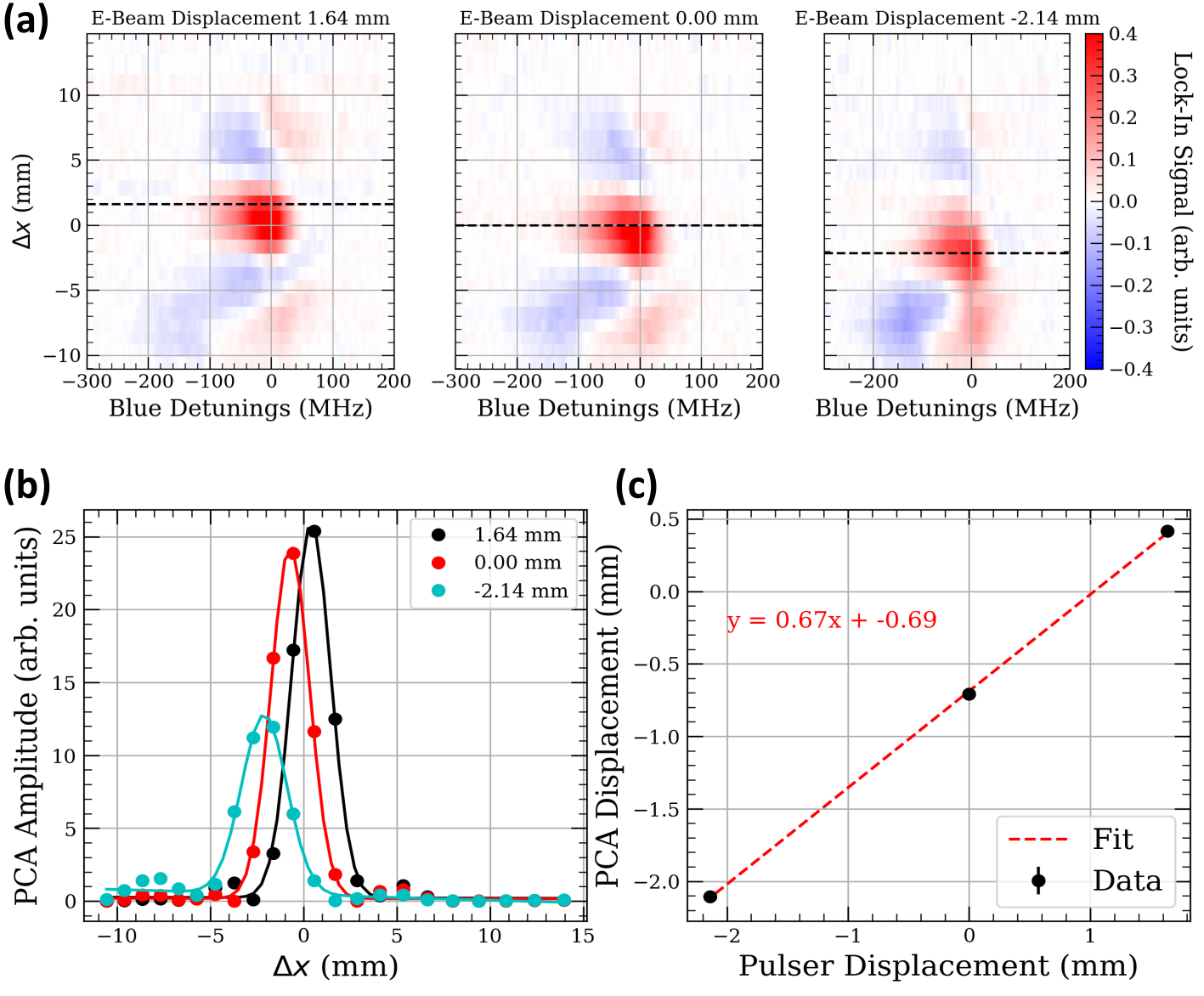}
    \caption{Measurements of the \textit{e}-beam motion along $x$ direction.
    (a) Heat maps of lock-in signal spectra vs the horizontal blue laser raster displacement $\Delta x$ for different horizontal positions
    of the \textit{e}-beam.
    The black dashed lines show the expected centers of the electron beam provided by the pulser deflection.
    (b) Calculated PCA amplitudes of the heat maps in (a) vs the horizontal laser beam displacement and their fits with a Gaussian shape.
    (c) The Gaussian  center location of fits in (b) vs the pulser dialed displacement.
    The red dashed line is a linear fit line showing relationship between measured and dialed \textit{e}-beam positions.
	For this dataset the \textit{e}-beam current was set to 108~$\mu$A.
    \label{fig:horizontal_displacement}
    }
\end{figure}

To confirm that we are sensitive to the $x$ direction motion of the \textit{e}-beam, we displace the \textit{e}-beam horizontally with the pulser deflection controls
and observed that the lock-in spectra maximum also moves in sync with the horizontal position sweep (see Fig.~\ref{fig:horizontal_displacement}).
Fig.~\ref{fig:horizontal_displacement}(a) is the lock-in spectra heat map vs the blue laser horizontal ($x$-direction) position scan and
it shows that the maximum of the signal is moving proportionally to the \textit{e}-beam displacement. We show the spectra
center position vs the \textit{e}-beam displacement in Fig.~\ref{fig:horizontal_displacement}(c), there is a clear linear relationship but
we see that the measured beam center position does not follow one to one relationship. We attribute it to the charging effects in the chamber which skew the relationship. However, in a real setup it can be calibrated in advance and still provides a useful \textit{e}-beam position monitor.

\begin{figure}
    \includegraphics[width=1\linewidth]{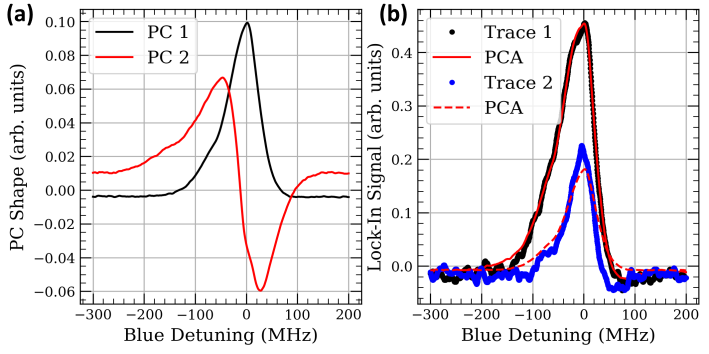}
    \caption{PCA analysis of the experimental data.
    (a) First two principle components (PC) or shapes that describe the data.
    (b) Example of two lock-in spectrum traces and their approximation with the weighted sum of these two principal components.
    \label{fig:pca_explain}
    }
\end{figure}

\section{Extraction of the e-beam parameters}

Since every raster point in $x-y$-plane acquires the lock-in signal spectrum, we need to perform data reduction to generate
a metric related to the \textit{e}-beam strength at this position. We chose to use 
principle component analysis (PCA) to extract a physically relevant parameter. The PCA 
extracts characteristic patterns (principal components) which describe the experimental data~\cite{brunton2022data}
(i.e. the lock-in signal spectra vs the blue laser detuning). Theoretically the number of needed patterns is equal to number of
collected spectra, but PCA sorts the patterns in the order of their contribution. Typically only a couple of first patterns are
required to provide a good description for the data (see Fig.~\ref{fig:pca_explain} where only two patterns were used). The contribution
of the other patterns monotonically drops and they could be thought as  describing the noise in the data or its outliers. Consequently,
every lock-in spectrum $S_i \approx w_{1_i} \times \textrm{PC}_1 + w_{2_i} \times \textrm{PC}_2$, where $i$ is index of the measured spectrum, $\textrm{PC}_i$ is the shape of the principal component, and $w_{1,2_i}$ is the weight (contribution) of corresponding component.
We chose to use the amplitude $A_i=\sqrt{w_{1_i}^2 + w_{2_i}^2}$ as a metric describing the \textit{e}-beam strength at the given
raster point (we refer to it as the PCA amplitude).

\begin{figure}
    \includegraphics[width=1\linewidth]{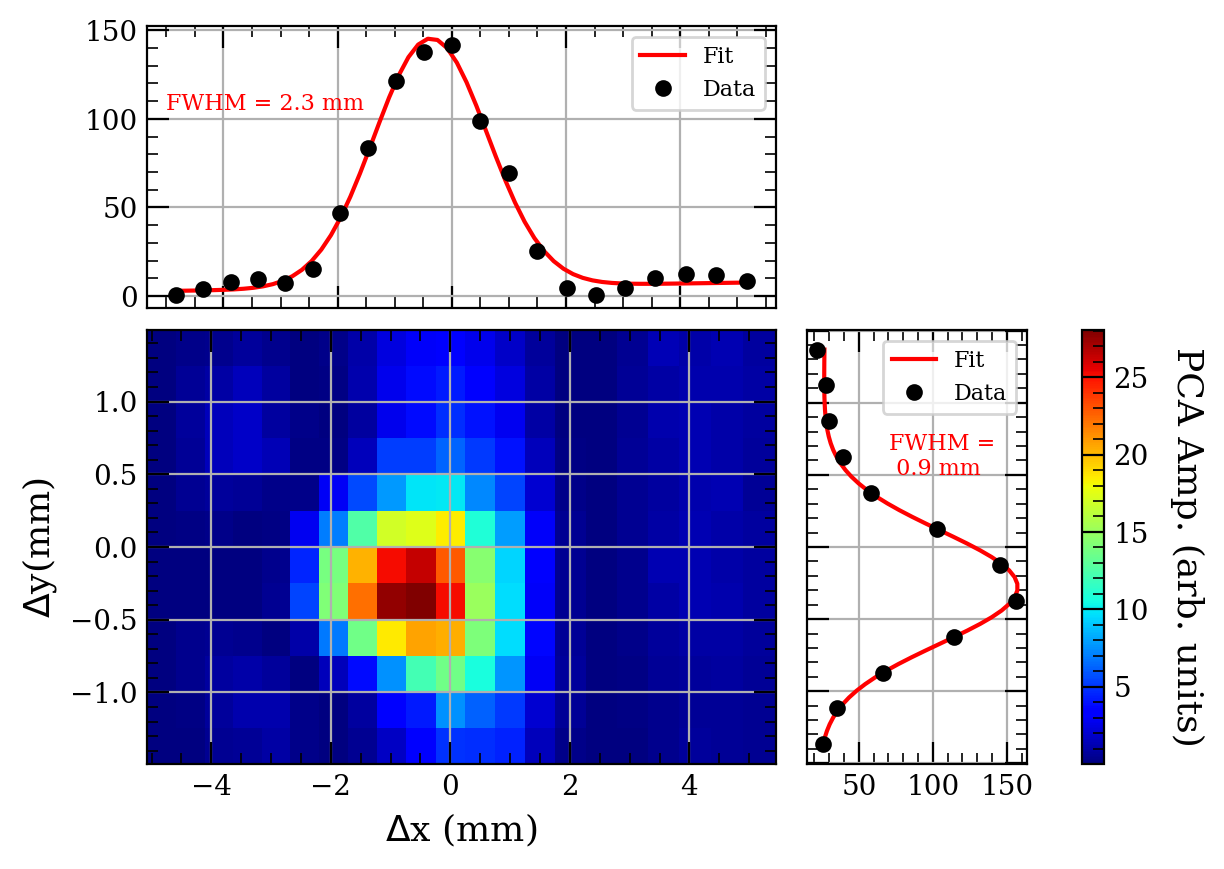}
    \caption{The PCA amplitude raster map of the \textit{e}-beam raster.
	The \textit{e}-beam current was set to 108~$\mu$A.
    \label{fig:beam_profile}
    }
\end{figure}

\begin{figure}
    \includegraphics[width=1\linewidth]{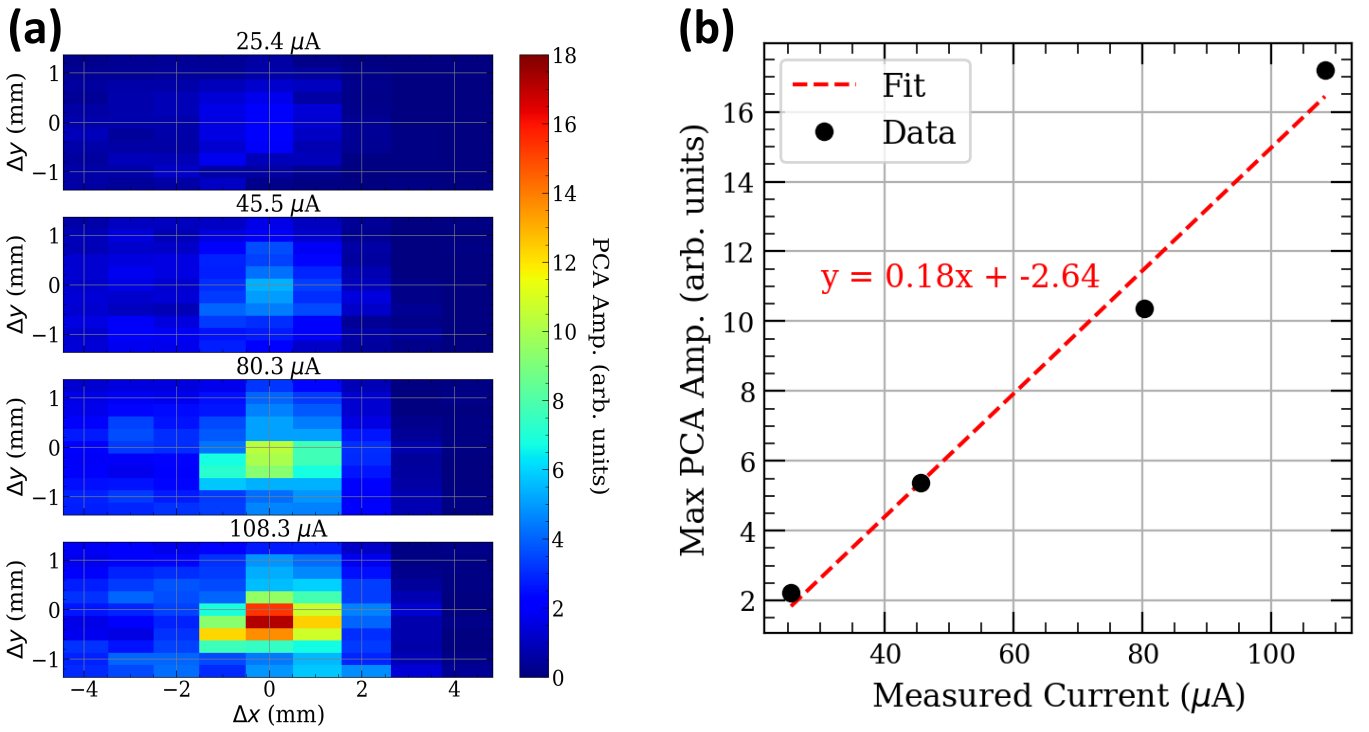}
    \caption{
    (a) The PCA amplitude raster maps for different \textit{e}-beam currents. The displacement steps are not as fine as those in Fig.~\ref{fig:beam_profile}.
    (b) Dependence of the PCA amplitude maximum in the raster map vs the \textit{e}-beam current for rasters in (a). The dashed red line is a linear fit.
    \label{fig:current_calibration}
    }
\end{figure}

We measured the \textit{e}-beam $x-y$ raster (profile) and show resulting heat
map of the PCA amplitude as a function of displacement relative to the
\textit{e}-beam in Fig.~\ref{fig:beam_profile} and Fig.~\ref{fig:current_calibration}(a).
The \textit{e}-beam appears stretched along $x$-direction but this is to be expected since the spatial response is determined
by the largest size from the actual \textit{e}-beam cross sections and the laser beam overlapping region. For the $x$-direction
we estimated the laser beam overlap size as $5$~mm. The FWHM size in this direction is about 2.3~mm which is smaller than
the overlap size, but a more precise estimate would require calculation of the beams shapes and the EIT power response  convolution,
which is smaller than our simple estimate. The vertical ($y$-direction)
FWHM of 0.9~mm is limited by the \textit{e}-beam size, size since it is limited by the width of the lasers.

We found that the maximum PCA amplitude in rastered 2D profiles has linear dependence on the \textit{e}-beam current as shown Fig.~\ref{fig:current_calibration}(b). This dependence can be used as a calibration curve in a non-invasive \textit{e}-beam current monitors.

\section{Conclusions}

This paper shows how one can use Rydberg atoms as non-invasive sensors to generate a 2D profile of
an \textit{e}-beam to deduce its size, position, and current using only optical transmission measurements. This technique can easily be extended for a 3D profiling. The resolution
of the profile is limited by the largest limiting dimension of the overlap
between the laser beams and the \textit{e}-beam. 
In contrast to the Rydberg fluorescence profiling~\cite{SchlossbergerOL25,BeharyPRR2025}, the presented laser beam crossing 
techniques can be used where an extra optical port is unavailable or in environments which are too harsh to collect images on a camera due to presence of, e.g., X-rays or ionizing radiation. The presented method is not limited to only \textit{e}-beam
detection but can be used for characterization any spatial distribution of the electric fields, for example generated near plasma
boundaries.

\section{Acknowledgments}
This work is supported by U.S. DOE Contract DE-SC0024621 and DE-AC05-06OR23177, NSF award 2326736 and Jefferson Lab LDRD program.

\bibliographystyle{spiebib} 
\bibliography{bibliography}
\end{document}